\documentclass[aps,prl,superscriptaddress,showpacs,twocolumn,floatfix,amsmath,amssymb]{revtex4-2}
\usepackage{hyperref} 
\usepackage[mathlines]{lineno} 
\usepackage{orcidlink}

\usepackage{graphicx} 
\usepackage{xcolor}
\usepackage[version=4]{mhchem}
\usepackage{graphicx}
\usepackage{xfrac}

\def\Fn{F_{2n}}
\def\Fp{F_{2p}}
\def\Fd{F_{2d}}
\def\Fh{F_{2h}}
\def\Ft{F_{2t}}
\def\FA{F_{2A}}

\def\Fpbar{\overline F_{2p}}
\def\Fnbar{\overline F_{2n}}

\def\Rh{R_h}
\def\Rt{R_t}

\def\dprat{\ensuremath{F_{2d}/F_{2p}}}
\def\nprat{\ensuremath{F_{2n}/F_{2p}}}
\def\htrat{\ensuremath{F_{2h}/F_{2t}}}
\def\Rht{\ensuremath{\mathcal{R}_{ht}}}
\def\Rth{\ensuremath{\mathcal{R}_{th}}}

\def\gtorder{\mathrel{\raise.3ex\hbox{$>$}\mkern-14mu
 \lower0.6ex\hbox{$\sim$}}}
\def\ltorder{\mathrel{\raise.3ex\hbox{$<$}\mkern-14mu
 \lower0.6ex\hbox{$\sim$}}}

\begin{document}
\title{Impact of the MARATHON data on \texorpdfstring{$F_{2n}/F_{2p}$}{F2n/F2p} and off-shell effects in light nuclei}


\author{T.~J.~Hague\orcidlink{https://orcid.org/0000-0003-1288-4045}}
\affiliation{Lawrence Berkeley National Laboratory, Berkeley, California 94720, USA}

\author{J.~Arrington\orcidlink{https://orcid.org/0000-0002-0702-1328}}
\affiliation{Lawrence Berkeley National Laboratory, Berkeley, California 94720, USA}

\author{S.~Li\orcidlink{https://orcid.org/0000-0003-1252-5392}}
\affiliation{Lawrence Berkeley National Laboratory, Berkeley, California 94720, USA}

\author{S.~N.~Santiesteban\orcidlink{https://orcid.org/0000-0001-5920-6546}}
\affiliation{University of New Hampshire, Durham, New Hampshire 03824, USA}

\date{\today}

\begin{abstract}

The neutron structure function, $\Fn$, has historically been extracted from measurements of the deuteron structure function, but our understanding of the nuclear effects on the bound proton and neutron limits the extraction of $\Fn$. 
The MARATHON collaboration recently extracted $\Fn$ from the comparison of $^3$H and $^3$He targets, where the nuclear effects are larger but nearly identical, yielding a precise extraction of $\nprat$.
This precise extraction can then be compared to deuteron extractions, providing important constraints on the nuclear effects in the deuteron.  
To ensure that this comparison is not biased by the specific model of nuclear effects used by MARATHON, we examine a range of models of the nuclear effects to obtain a more conservative, but more model-independent, extraction of $\nprat$ for comparison with deuteron extractions. 
Even with the more conservative approach, the comparison suggests the need for significant off-shell corrections or other nuclear effects, beyond those include in most calculations, even for the weakly-bound deuteron.

\end{abstract}

\maketitle


Deep inelastic scattering (DIS) from nucleons allows extraction of their structure functions which provide access to their quark parton distribution functions (pdfs). These pdfs depend on the quark longitudinal momentum fraction, $x$, and the four-momentum transfer of the probe, $Q^2$. A comparison of the proton and neutron structure functions, $\Fp$ and $\Fn$, allows us to study the difference between the up- and down-quark contributions to the nucleon pdfs, since the proton represents a bound (uud) state and the neutron a (udd) state. While $\Fp$ is precisely measured over a wide kinematic range, $\Fn$ is less well known, especially at large values of Bjorken-$x$ where the struck quark carries a large fraction of the neutron's longitudinal momentum. The lack of a free neutron target makes measurements of $\Fn$ more challenging, and most methods of isolating scattering from the neutron introduce enhanced experimental uncertainties and model dependence in the extraction.

For many years, the most common way to extract $\Fn$ was to measure the deuteron structure function, $\Fd$, subtract the proton contribution, and apply a model-dependent correction for the modification of the free nucleon structure functions in the deuteron due to binding, Fermi motion, and other effects~\cite{Arrington:2008zh, Accardi:2009br, Li:2023yda}. Because $\Fp > \Fn$, especially at large $x$, the subtraction introduces a large enhancement in the experimental uncertainties. The model dependence of the nuclear corrections yields additional uncertainties that dominate at large $x$ values. As such, the $\Fn$ extractions have larger uncertainties than $\Fp$, and these uncertainties grow with $x$, especially above $x \approx 0.6$~\cite{Accardi:2011fa, Arrington:2011qt}

An alternative approach was proposed~\cite{Afnan:2000uh, Afnan:2003vh} to compare the helion ($h$, $^3$He nucleus) and triton ($t$, $^3$H nucleus). These $A=3$ mirror nuclei differ in that the number of protons and neutrons are swapped, and the expectation is that their structure should be nearly identical up to small differences associated with, for example, the isospin-symmetry breaking effects associated with the Coulomb interaction~\cite{wiringa14}. While the binding, smearing, and off-shell effects will be larger in these $A=3$ nuclei than in deuterium, the nuclear corrections should be nearly identical for the helion and triton, yielding a much smaller model dependence in the extraction of $\nprat$~\cite{Afnan:2000uh, Afnan:2003vh} from the $\htrat$ ratio.

The MARATHON experiment~\cite{MARATHON:2021vqu} provided precision measurements of $\htrat$ and extracted $\nprat$ using the Kulagin-Petti (KP) model~\cite{Kulagin:2004ie,Kulagin:2010gd} to evaluate the size and uncertainties of the nuclear effects. The relative normalization of the $\Fh$ and $\Ft$ results was determined based on a comparison of the values of $\nprat$ extracted from $h/t$ and $d/p$ measurements, which depends on the nuclear effects in both cases. To avoid questions about whether this procedure may bias the comparison, we extract $\nprat$ using a range of nuclear models as a more conservative evaluation of the model dependence. We also take a more conservative approach to normalizing the $h/t$ ratios. We then compare this more conservative, but less model dependent, result to extractions from worlds $\dprat$ data to determine the extent to which we can use this to constrain nuclear effects in the deuteron. 

Note that extracting pdfs, rather than simply structure functions, requires disentangling additional effects: higher-twist contributions, target mass corrections, and effects beyond leading order. In this work we compare $h/t$ and $d/p$ extractions of $\nprat$ to assess structure function level conclusions while avoiding these additional complications. In the end, we also compare to pdf analyses that are addressing similar questions using both the structure function ratios and a variety of other data sets~\cite{Cocuzza:2021rfn}.


We begin by examining nuclear effects in the extraction of $\nprat$ from the measurements on light nuclei. In general, we define the nuclear corrections to be the ratio of the nuclear structure function to the sum of its constituent nucleons, $R_A = \FA / (Z \Fp + N \Fn)$. This includes both modification of the nucleon structure function for bound nucleons, and any other contributions that modify the nuclear structure functions. 

Calculations of the deuteron structure function typically include the proton and neutron contributions, modified by smearing and binding, plus additional terms. Following Ref.~\cite{Arrington:2011qt}, we take $\Fd = \Fpbar + \Fnbar + \delta \Fd$ where $\overline F_{2N} = S_N F_{2N}$ is be the nucleon structure function contribution after smearing and binding, and $\delta \Fd$ represents additional contributions. While $S_N$ depends on both the convolution formalism and the free nucleon structure function, for a given calculation we can express it as a smearing function $S_N(x)$. We can then absorb the $\delta \Fd$ contribution into the smearing function by taking $\Delta = \delta \Fd / \Fd$ and modify the smearing correction as $\widetilde{S}_N = S_N / (1-\Delta)$.  This allows us to express the deuteron structure function as $\Fd = \widetilde{S}_p \Fp + \widetilde{S}_n \Fn$, and yields the following expression for $\nprat$ as a function of $\dprat$:
\begin{equation}
\frac{F_{2n}}{F_{2p}}
= {\frac{1}{\widetilde{S}_n}}
  \left( \frac{F_{2d}}{F_{2p}} - \widetilde{S}_p \right).
\label{eq:np-from-dp}
\end{equation}
Note that this equation corrects an error in eq. (1) of Ref.~\cite{Arrington:2011qt}, which included a factor of $\Fp$ in the final term, which appeared in the equation but was not included in their analysis. The factor $\widetilde{S}_N$ will depend on the convolution model, off-shell effects, and input structure functions. Ref.~\cite{Arrington:2011qt} evaluated $\widetilde{S}_N$ for a range of different inputs to examine the sensitivity of $\nprat$ to the nuclear effects as a function of $x$.

\begin{figure}
    \centering
    \includegraphics[width=0.46\textwidth]{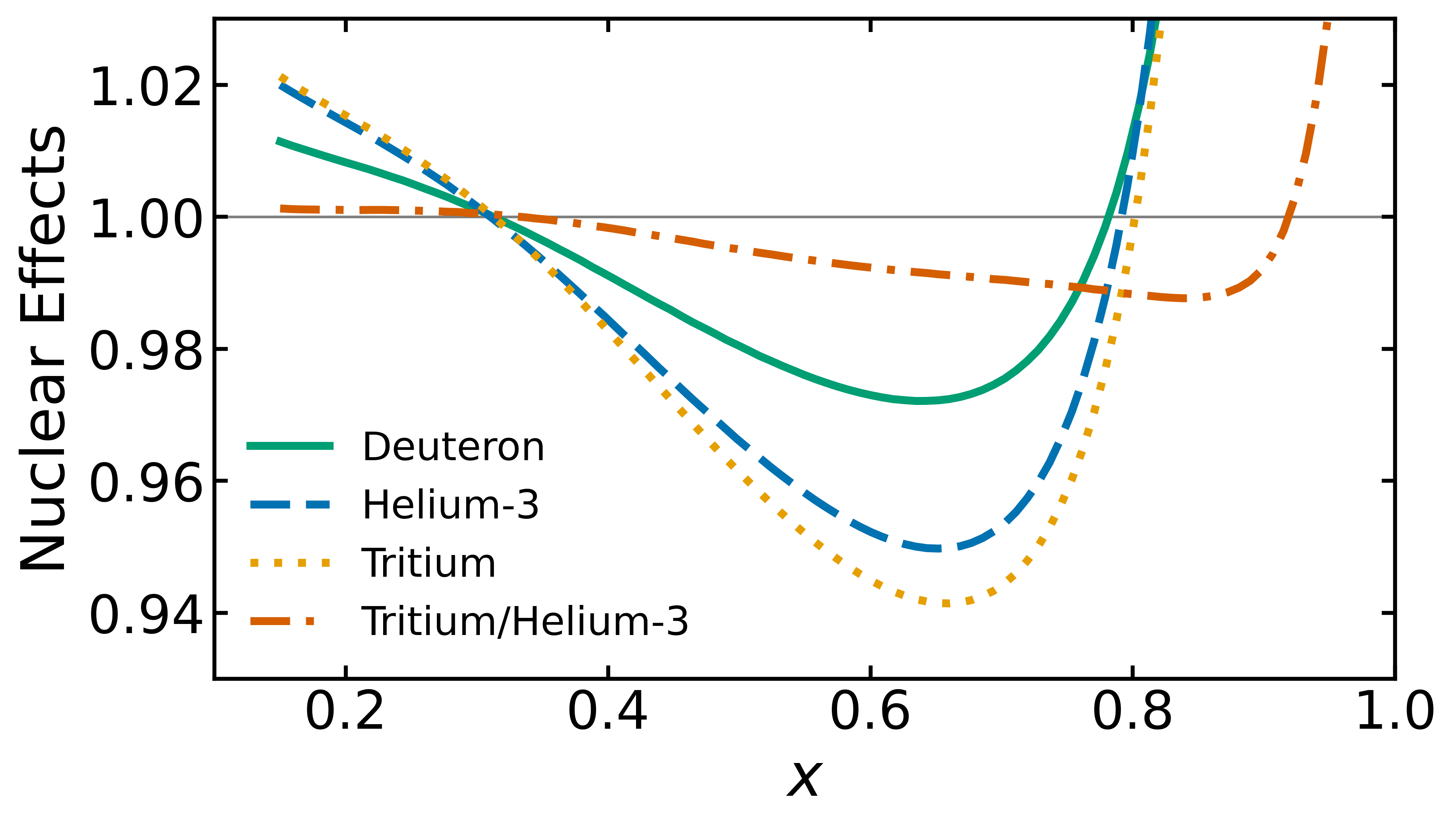}
    \caption{The nuclear effects, $\FA / (Z \Fp + N \Fn)$ from the KP model~\cite{Kulagin:2010gd} for $^2$H (solid line), $^3$He (dashed line), $^3$H(dotted line), and the correction on the $t/h$ ratio $\Rth$ (dash-dotted line). Note that the nuclear effects for the $h/t$ extraction are smaller than for the $d/p$ case, and the larger effects and rapid $x$ dependence occur at larger $x$ values.}
    \label{fig:nuclear_effects}
\end{figure}

In the MARATHON experiment~\cite{MARATHON:2021vqu}, $\nprat$ was extracted using new and precise measurements of the $\htrat$ ratio for $0.195 < x < 0.825$. In this case, the nuclear effects are written out in terms of their impact on the helion and triton, rather than in terms of modified nucleons, with $\Rh = \Fh / (2\Fp + \Fn)$ and $\Rt = \Ft / (\Fp + 2 \Fn)$. Taking $\Rht = \Rh / \Rt$, we can extract $\nprat$ from $\htrat$ and a calculation of $\Rht$:
\begin{equation}
\frac{F_{2n}}{F_{2p}} =
\frac{2 \Rht - \htrat}{2 \htrat - \Rht}~.
\label{eq:np-from-ht}
\end{equation}

Figure~\ref{fig:nuclear_effects} shows the size of the nuclear effects for $^2$H, $^3$H, $^3$He, and in the $^3$H/$^3$He ratio~\cite{Kulagin:2010gd}. For the $A=3$ nuclei, the nuclear effects are significantly larger than for the deuteron extraction, as expected. Because the extraction from the $h/t$ ratio depends only on the ratio of nuclear effects, $\Rht$, the correction is significantly smaller than either case on their own. In addition, the rapid change in the $x$ dependence, occurring at $x\approx0.8$ for the deuteron or $A=3$ nuclei is pushed out to $x>0.9$ for $\Rht$. This also yields much smaller uncertainties in the nuclear effects, as examined in Ref.~\cite{Afnan:2003vh}, which examined a variety of models and concluded that the corrections were small with uncertainties at the $\ltorder$1\% level to $x=0.8$. The MARATHON extraction~\cite{MARATHON:2021vqu} used the KP model~\cite{Kulagin:2004ie, Kulagin:2010gd} along with the parton distributions of~\cite{Alekhin:2017fpf} to calculate $\Rht$. Again, the nuclear effects were found to be small in the ratio, as were the estimated uncertainties for this model: $<$0.1\% up to $x=0.65$, growing to 0.43\% at $x=0.825$.


\begin{figure}
    \centering
    \includegraphics[width=0.46\textwidth]{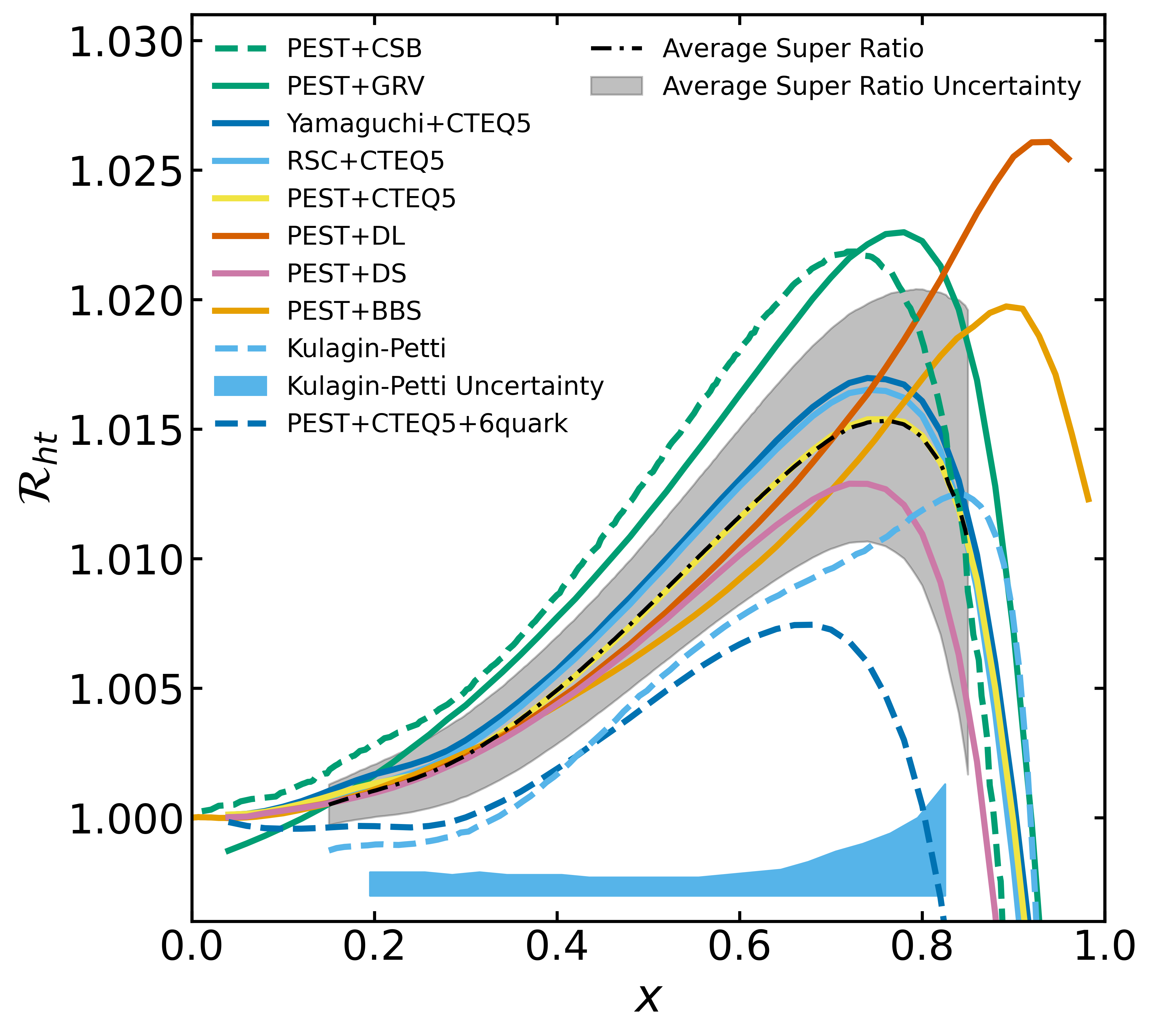}
    \caption{Selected calculations of $\Rht$ from Ref.~\cite{Afnan:2003vh} alongside the KP model~\cite{Kulagin:2004ie,Kulagin:2010gd} that was used by the MARATHON experiment. Also shown is the average of all plotted models with a $1\sigma$-rms band with a dash-dotted line. Models that include effects beyond smearing and off-shell effects are shown with dashed lines. The legend is ordered by descending values of the model evaluated at $x=0.6$. The blue band shows the uncertainty from the KP model used in Ref.~\cite{MARATHON:2021vqu}.}
    \label{fig:all_super}
\end{figure}

We repeat the MARATHON extraction of $\nprat$ by examining a range of models of $\Rht$ to assess the model dependence of the extraction. For this study, we have chosen to use a subset of the $\Rht$ calculations from Ref.~\cite{Afnan:2003vh} plus the KP model as used by the MARATHON experiment. The calculations in this reference vary by technique for calculating wave functions, choice of nuclear potential, choice of nucleon pdfs, and the inclusion of more exotic effects. The analysis presented here considers calculations that use wave functions calculated via the Faddeev equations, as they involve fewer approximations in the calculations. To best capture the variance of the calculations, each nuclear potential is paired with the CTEQ5 pdfs and then each pdf set is paired with the PEST potential. These selected calculations are shown in Fig.~\ref{fig:all_super} alongside the KP model. Also shown is the average of the calculations and the $1\sigma$ root mean square (rms) uncertainty band that we use to quantify the model uncertainty of the $\Rht$ calculations. In figure~\ref{fig:all_super}, dashed lines are used to indicate calculations that introduce effects beyond smearing and off-shell effects, e.g. contributions from six-quark bags, charge-symmetry violation, or higher-twist contributions that do not cancel in the $h/t$ ratios. Not surprisingly, these increase the rms scatter of the models, and to be conservative we take the full set of models shown in evaluating the uncertainty band used in our analysis. If we were to focus only on more conventional effects, the rms of the distribution would be lower by 30--60\% over MARATHON data range. 

The MARATHON collaboration found that their $d/p$ cross section ratios were consistent with previous SLAC~\cite{Bodek:1979rx} data. In light of this agreement, a (2.5$\pm$0.7)\% renormalization was applied to the $h/t$ ratios so that the $\nprat$ values extracted from were in agreement with the values extracted from $d/p$ at $x=0.31$. This procedure has the drawback that it relies on the KP model for both the $h/t$ and the $d/p$ nuclear effects. One might be concerned that this goes against the idea of minimizing the model dependence by avoiding corrections to the deuteron, although the corrections for deuterium at low $x$ have smaller uncertainties than they do at larger $x$ values.

While MARATHON found that a (2.5$\pm$0.7)\% change in the nominal $h/t$ target thickness ratio was needed, this is somewhat larger than the nominal 1.1\% normalization uncertainty from the measurement of the target thicknesses~\cite{Li:2022fhh, Arrington:2023hht, Santiesteban:2023rsh}. A separate global pdf analysis~\cite{Cocuzza:2021rfn} also found that a much smaller correction of (0.7$\pm$0.6)\% provided the best consistency with world's data given their model of the nuclear effects. These disagreements indicate that the normalization procedure has a significant model dependence. In this work, we take a conservative approach by applying a renormalization correction of (1.25$\pm$1.25)\%. This is large enough that the one-$\sigma$ region includes both the nominal target thickness measurements and the MARATHON renormalization adjustment. This approach increases the total uncertainty, but does not significantly limit the sensitivity to nuclear effects in the comparison of $h/t$ and $d/p$ measurements. 

\begin{figure}[ht]
    \includegraphics[width=0.46\textwidth]{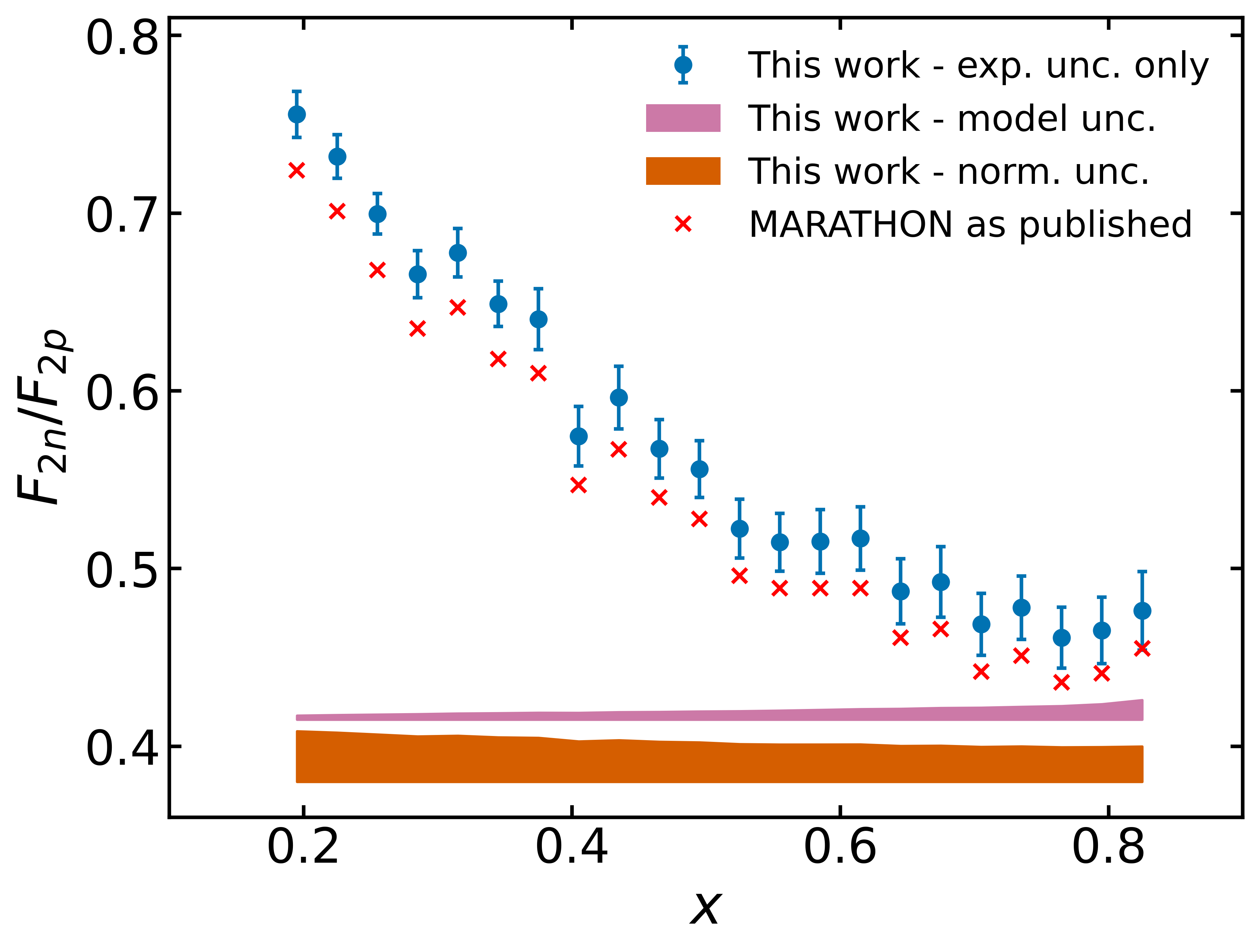}
    \caption{$\nprat$ and uncertainties from this analysis, compared to the values extracted from the original extraction~\cite{MARATHON:2021vqu}. The error bars on the values from this analysis include both statistical and point-to-point systematic uncertainties. The model uncertainty is the impact of the 1$\sigma$ uncertainty on $\Rht$ and the normalization band shows the correlated shift associated with the $1.25\%$ normalization uncertainty in $\htrat$.}
    \label{fig:np-modeldep}
\end{figure}

Using the average value of the $\Rht$ calculations in Fig.~\ref{fig:all_super} and applying the 1.25\% normalization factor to the measured $h/t$ ratio yields a modified extraction of $\nprat$, shown in Fig.~\ref{fig:np-modeldep}. In this figure, the points include the experimental uncertainties, while the bands indicate the model dependence, calculated using the 1$\sigma$ rms scatter of $\Rht$ in Fig.~\ref{fig:all_super}, and renormalization uncertainties described above. Despite the large differences between the $\Rht$ calculations, it is clear that the uncertainties are dominated by the experimental data and the normalization uncertainties. While the $h/t$ normalization factor has a significant contribution to the final uncertainty and modifies the value of the extracted $\nprat$, it has minimal impact on its shape. Compared to the original MARATHON result, it increases $\nprat$ by about 0.025 with a slightly smaller effect at large $x$. Because this result shifts up but has a slightly faster falloff with $x$, the effect of a change in the $h/t$ normalization has a reduced impact on extrapolation of the curve to $x=1$~\cite{Cui:2021gzg, Valenty:2022gqm}.

\begin{figure}[htb]
  \includegraphics[width=0.46\textwidth,trim={0mm 0mm 0mm 0mm}, clip]{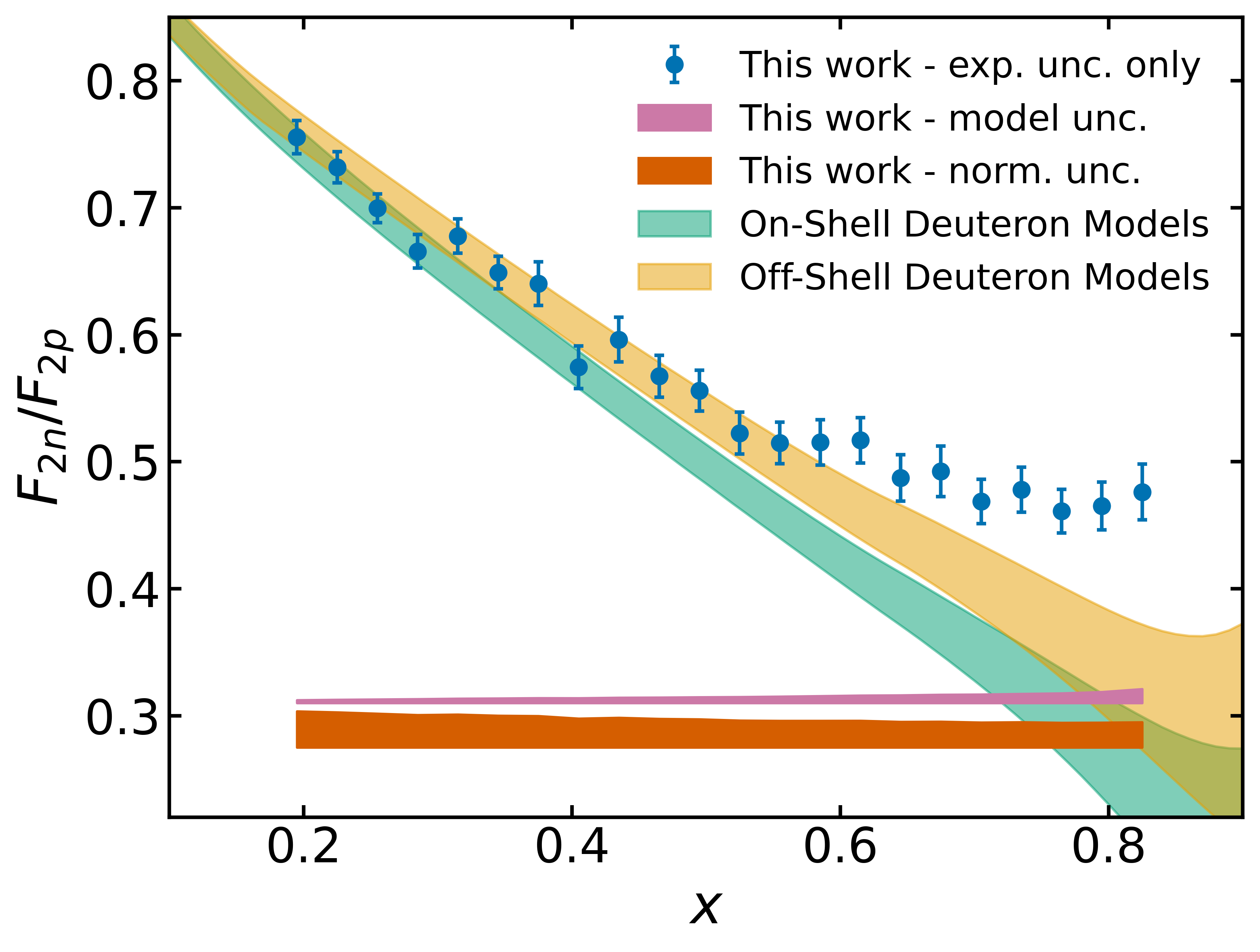}
  \caption{Comparison of this work (circles) and $\nprat$ extractions from $d/p$ measurements with (top band) and without (bottom band) the inclusion of off-shell effects~\cite{Arrington:2011qt}.}
\label{fig:ht-vs-dp}
\end{figure}

Even with the more conservative estimation of model dependence, this extraction of $\nprat$ has significantly smaller uncertainty than the extractions from $d/p$ measurement at large $x$. Thus, these results can be used to constrain the nuclear corrections to the $d/p$ measurements.  Figure~\ref{fig:ht-vs-dp} shows $\nprat$ from this analysis alongside the range extracted from $d/p$ extractions using models, with and without off-shell effects, from Ref.~\cite{Arrington:2011qt}. Enhanced deuteron off-shell effects cause a rise in the extracted $\nprat$ value, with the highest edge of this range corresponding to larger off-shell effects.

The $A=3$ results are in good agreement with $d/p$ extractions at low $x$, and the assumed uncertainty on the normalization factor is sufficient to cover the full range of deuteron extractions in this region. The extracted $\nprat$ values show a significant difference at larger $x$ values, where the nuclear effects are large. Note that while an increase in the $h/t$ normalization factor could reconcile the extractions for $x>0.6$, it would require a 4$\sigma$ shift, even with our more conservative uncertainty estimates, to be consistent with the upper edge of the deuteron off-shell extractions. This would resolve the high-$x$ discrepancy but would reduce $\nprat$ by almost 0.1 in the low-$x$ region, causing a more significant disagreement with the $d/p$ extractions in the region where the nuclear uncertainties are believed to be smallest.

This discrepancy suggests that nuclear effects, beyond those included in the calculations examined here, are needed to yield consistent extractions. The simplest explanation would be larger off-shell effects in the deuteron, which would increase the extracted $\nprat$ above the top of the yellow band in Fig.~\ref{fig:ht-vs-dp}. However, it is also possible that some more exotic effect significantly modifies the $d/p$ extraction or even the $h/t$ extraction at large x. One such possibility is the inclusion of isospin-dependent nuclear corrections~\cite{Arrington:2015wja, Arrington:2019wky}, which are not included in any of the models examined here.

Our analysis shows a clear discrepancy between the extractions from $d/p$ and $h/t$ cross section ratios, based on a wide range of models for the nuclear effects. The benefit of working at the structure function level is that our conclusions depend only on the consistency of the structure function data and associated nuclear corrections. This makes the interpretation more straightforward than global pdf analyses, which are sensitive to tension between a range of different experimental data sets as well as the theory connecting the pdfs to more complicated observables. The limitation is that we do not separate the structure function into contributions from the pdfs and other QCD effects such as target mass corrections~\cite{Nachtmann:1973mr, PhysRevD.14.1829} and higher twist effects~\cite{Accardi:2009br}, as in a global pdf analyses~\cite{Cocuzza:2021rfn, PhysRevD.107.L051506, Li:2023yda}. Because of this, we cannot probe the interplay between isospin-dependent off-shell effects and higher-twist effects, suggested to be important in the CJ global QCD analysis~\cite{cj_ht}.

Following the release of MARATHON results, efforts have been made to reconcile the $\nprat$ extraction from deuteron and $A=3$ nuclei by including parameterized flavor-dependent off-shell and/or higher-twist effects into a global pdf analysis. Such an analysis from the JAM collaboration~\cite{Cocuzza:2021rfn} claimed a large flavor-dependent off-shell effect, while a later study~\cite{PhysRevD.107.L051506} reported that, with their additive higher-twist parameterization that described the MARATHON data well, the observed off-shell function difference between proton and neutron is consistent with zero. Recently, the CJ collaboration updated their $n/p$ extractions from world proton and deuteron data by including new and more precise measurements from Jefferson Lab as well as a detailed study on uncertainties~\cite{Li:2023yda}. The result is consistent with the off-shell deuteron models in Fig.~\ref{fig:ht-vs-dp} and still deviates from the $A=3$ extractions at large $x$. The fact that the conclusions depend on assumptions made in the analysis indicates that we still don't have sufficient knowledge to fully constrain the isospin dependence of nuclear effects, but demonstrate that the MARATHON data provide the first opportunity to perform significant studies of the isospin dependence. 

In conclusion, we have presented an updated extraction of $\nprat$ from the MARATHON data using a more conservative analysis aimed at minimizing the model dependence of the extracted ratio. While the uncertainties are larger than the original analysis~\cite{MARATHON:2021vqu}, they are still a significant improvement over extractions based on $d/p$ measurements. Our results are in good agreement with extractions based on $d/p$~\cite{Arrington:2011qt} up to $x \approx 0.6$, while at larger $x$ values, our results are significantly higher.

Because of the larger model dependence in the extraction of $\nprat$ from $d/p$ measurements, it is natural to conclude that this difference indicates the presence of larger nuclear effects for the deuteron; either larger off-shell corrections or other effects beyond binding and Fermi motion.  Another possibility~\cite{Cocuzza:2021rfn} is that isospin-dependent off-shell or higher-twist effects change the $h/t$ extraction while cancelling, at least in part, in the deuteron extraction. In this case, it is possible to have cancellation between proton and neutron corrections such that the impact is larger in the $h/t$ extraction than for $d/p$, but the general conclusion remains: corrections beyond those included in most nuclear models appear to be necessary.

One such option is the contribution of isospin-dependent nuclear corrections which would imply a flavor-dependent EMC effect, a topic that has received significant attention recently. The non-trivial correlation between the nuclear effects~\cite{seely09} and the presence of short-range correlations (SRCs)~\cite{fomin12} in light nuclei, combined with dominance of np-SRCs~\cite{Arrington:2022sov} suggests a mechanism for isospin-dependent nuclear effects~\cite{arrington12b, Arrington:2015wja}. This has been examined in more detail in recent analyses~\cite{schmookler19, Arrington:2019wky}, demonstrating the the correlation has potential sensitivity to the question of flavor dependence, but not yet providing evidence for or against this. The sensitivity will be improved with the addition of EMC and SRC measurements for a range of new targets~\cite{e1206105, e1210108}, while parity-violating electron scattering in a non-isoscalar target such as $^{48}$Ca can provide a clean and precise measurement of such flavor dependence~\cite{Cloet:2012td, JeffersonLabSoLID:2022iod, Beminiwattha:2023est}.

Future JLab measurements will provide additional constraints on both the nuclear effects in the deuteron and on the possibility of isospin- or flavor-dependent nuclear corrections~\cite{Arrington:2021alx}. Measurements of scattering from the neutron in deuterium with a tagged spectator proton provide another way to extract $\nprat$ with reduced (and very different) model dependence~\cite{CLAS:2011qvj, CLAS:2014jvt}, and have also be used to extract nuclear effects in the deuteron~\cite{Griffioen:2015hxa}. The BoNuS12 experiment~\cite{bonus12} has made such tagged measurements in CLAS12, which should significantly improve the precision and the coverage at large $x$. Parity violating electron-proton DIS scattering in SoLID~\cite{JeffersonLabSoLID:2022iod, Beminiwattha:2023est} can provide a model-independent extraction by using the flavor sensitivity of the weak probe, rather than using nuclear targets, to isolate the $d/u$ quark ratio.

\begin{acknowledgments}

We thank Alberto Accardi, Wally Melnitchouk, and Nobuo Sato for useful discussions, and WM and NS for providing calculations from Ref.~\cite{Afnan:2003vh}. This work was supported by the Department of Energy's Office of Science, Office of Nuclear Physics, under contracts DE-AC02-05CH11231 and DE-SC0024665.

\end{acknowledgments}

\bibliographystyle{apsrev4-2}

\bibliography{MARATHON}

\clearpage
\newpage

\appendix

\section{Extended Descriptions of Figures}

To meet the access needs of people that use screen-readers to consume our work, we have included extended descriptions of the figures in this appendix.

\subsection{Figure 1}

The horizontal axis is labeled ``x'' and extends from 0.1 to 1.0. 
The vertical axis is labeled ``Nuclear Effects'' and extends from approximately 0.93 to 1.03. 
There is a horizontal line to note when Nuclear Effects are 1. A green, solid line is labeled ``Deuteron'' that is approximately linear from 1.01 at x=0.18, crossing 1 at x=0.31, to 0.97 at x=0.6 when it begins to rapidly rise and is out of the axis bounds just past x=0.8. 
A blue, dashed line is labeled ``Helium-3'' that is approximately linear from 1.02 at x=0.18, crossing 1 at x=0.31, to 0.95 at x=0.6 when it begins to rapidly rise and is out of the axis bounds just past x=0.8. 
A yellow, dotted line is labeled ``Tritium'' that is approximately linear from 1.02 at x=0.18, crossing 1 at x=0.31, to 0.94 at x=0.6 when it begins to rapidly rise and is out of the axis bounds just past x=0.8. 
An orange, dash-dotted line is labeled ``Tritium/Helium-3'' that is approximately horizontal with a value of 1 from x=0.18 to x=0.31, after which it is approximately linearly decreasing to a value of 0.99 at x=0.9, and then begins to rapidly rise and is out of the axis bounds at x=0.95. 
This shows that the Nuclear Effects in Tritium and Helium-3 approximately cancel and pushes the less constrained region, when it rapidly rises, past the range of available data.

\subsection{Figure 2}

The horizontal axis is labeled ``x'' and extends from 0 to 1. 
The vertical axis is labeled ``R subscript ht'', with the R in a script font, and extends from approximately 0.995 to 1.031. 
A set of ten curves that show different calculations of the ratio of nuclear effects in helium-3 to tritium, as well as a curve showing the average of the ten calculations and the 1-sigma rms spread as a band around the curve. 
The curves are in somewhat close agreement at low x, with a +/- 0.002 spread but rapidly diverge with increasing x. 
The curve labels are ordered by descending value at x=0.6, which is: ``PEST+CSB'', ``PEST+GRV'', ``Yamaguchi+CTEQ5'', ``RSC+CTEQ5'', ``PEST+CTEQ5'', ``PEST+DL'', ``PEST+DS'', ``PEST+BBS'', ``Kulagin-Petti'', and ``PEST+CTEQ5+6quark''. 
All of the curves exhibit a rise above unity, most often beginning between x=0-0.2; the exceptions to this are the ``Kulagin-Petti'' and ``PEST+CTEQ5+6quark'' curves which both rise above unit around x=0.3. 
The curves rise and peak, before a rapid downturn with the peak occurring just below x=0.8; the exceptions are the ``Kulagin-Petti'', ``PEST+DL'', and ``PEST+BBS'' curves which peak closer to x=0.9 and the ``PEST+CTEQ5+6quark'' curve which peaks closer to x=0.7. 
The average curve begins at 1 at x=0.195, peaks just below x=0.8 at 1.015, then rapidly downturns to 1.010 at x=0.83. 
The rms band increases approximately linearly from +/- 0.001 at x=0.195 to nearly +/- 0.010 at x=0.83. 
A data table with the average value and the uncertainty evaluated at the MARATHON data points will be available in the final publication.

\subsection{Figure 3}

The horizontal axis is labeled ``x'' and extends from about 0.1 to 0.9. 
The vertical axis is labeled ``F2n/F2p'' and extends from about 0.35 to 0.81. 
Blue, circular points are labeled ``This work - exp. unc. only'' and decrease approximately linearly from 0.756 at x=0.195 to 0.469 at x=0.705 and then becomes approximately horizontal. 
The error bars increase approximately linearly from 0.013 to 0.022. 
A pink, solid band is labeled ``This work - model unc.'' and increases in width approximately linearly from 0.002 to 0.011. 
An orange, solid band is labeled ``This work - norm. unc.'' and decreases in width approximately linearly from 0.028 to 0.020. 
Red, ``x'' points are labeled ``MARATHON as published'' and decrease approximately linearly from about 0.72 at x=0.195 to about 0.45 at x=0.705 and then becomes approximately horizontal. 
A data table with the values of data labeled ``This work'' evaluated at the MARATHON data points will be available in the final publication. 
A data table for the ``MARATHON as published'' points can be found in Ref.~~\cite{MARATHON:2021vqu}.

\subsection{Figure 4}

The horizontal axis is labeled ``x'' and extends from about 0.1 to 0.9. 
The vertical axis is labeled ``F2n/F2p'' and extends from about 0.25 to 0.31. 
Blue, circular points are labeled ``This work - exp. unc. only'' and decrease approximately linearly from 0.756 at x=0.195 to 0.469 at x=0.705 and then becomes approximately horizontal. 
The error bars increase approximately linearly from 0.013 to 0.022. 
A pink, solid band is labeled ``This work - model unc.'' and increases in width approximately linearly from 0.002 to 0.011. 
An orange, solid band is labeled ``This work - norm. unc.'' and decreases in width approximately linearly from 0.028 to 0.020. 
A green, solid band is labeled ``On-Shell Deuteron Models'' and decreases approximately linearly from about 0.83 at x=0.1 to about 0.25 at x=0.85. 
The width of the band slowly increases from about 0.02 at x=0.1 to about 0.06 at x=0.8 after which is rapidly broadens. 
A yellow, solid band is labeled ``Off-Shell Deuteron Models'' and decreases approximately linearly from about 0.83 at x=0.1 to about 0.3 at x=0.9. 
The width of the band slowly increases from about 0.02 at x=0.1 to about 0.08 at x=0.8 after which is rapidly broadens. 
A data table with the values of data labeled ``This work'' evaluated at the MARATHON data points will be available in the final publication.

\end{document}